\documentclass[letterpaper]{article} 
\usepackage{times}  
\usepackage{helvet}  
\usepackage{courier}  
\usepackage[hyphens]{url}  
\usepackage{graphicx} 
\usepackage{caption} 
\usepackage{algorithm}
\usepackage{algorithmic}
\usepackage{fullpage}

\usepackage{newfloat}
\usepackage{listings}
\DeclareCaptionStyle{ruled}{labelfont=normalfont,labelsep=colon,strut=off} 
\floatstyle{ruled}
\newfloat{listing}{tb}{lst}{}
\floatname{listing}{Listing}
\usepackage{enumerate}
\usepackage{cite}
\usepackage{amsmath,amssymb,amsfonts}
\usepackage{textcomp}
\usepackage{mathtools}

\usepackage{hyperref}       
\usepackage{booktabs}       
\usepackage{bm}
\usepackage{xspace}
\usepackage{xcolor}
\usepackage{color, colortbl}
\usepackage{multirow}
\usepackage{threeparttable}

\hypersetup{
    colorlinks = true,
    linkcolor = [rgb]{0.6,0,0},
    citecolor = [rgb]{0,0.6,0},
    urlcolor = [rgb]{0,0,0.6}
}

\def\argmin{\mathop{\mathrm{arg\,min}}} 

\def\lim{\mathop{\mathrm{lim}}} 

\def\cbm{{\bm{c}}}

\def\sbm{{\bm{s}}}
\def\xbm{{\bm{x}}}

\def\varphibm{{\bm{\varphi}}}

\def\thetabm{{\bm{\theta}}}

\def\xbmhat{{\widehat{\bm{x}}}}

\def\thetabmhat{{\widehat{\bm{\theta}}}}

\def\R{\mathbb{R}}

\definecolor{maroon}{cmyk}{0,0.87,0.68,0.32}

\begin{document}

\title{SINCO: A Novel structural regularizer for image compression \\ using implicit neural representations}

\date{}

\author{
Harry Gao$^{*}$, Weijie Gan$^{*}$, Zhixin Sun, and Ulugbek S. Kamilov\\
\small $^*$These authors contributed equally to this work. \\
\small Computational Imaging Group, Washington University in St. Louis, MO 63130, USA\\
\small \texttt{\{harrygao, weijie.gan, zhixin.sun, kamilov\}@wustl.edu}
}

\maketitle

\begin{abstract}
    Implicit neural representations (INR) have been recently proposed as deep learning (DL) based solutions for image compression. An image can be compressed by training an INR model with fewer weights than the number of image pixels to map the coordinates of the image to corresponding pixel values. While traditional training approaches for INRs are based on enforcing pixel-wise image consistency, we propose to further improve image quality by using a new structural regularizer. We present \emph{\textbf{s}tructural regular\textbf{i}zatio\textbf{n} for INR \textbf{co}mpression (SINCO)} as a novel INR method for image compression. SINCO imposes structural consistency of the compressed images to the groundtruth by using a segmentation network to penalize the discrepancy of segmentation masks predicted from compressed images. We validate SINCO on brain MRI images by showing that it can achieve better performance than some recent INR methods.
\end{abstract}

\section{Introduction}
\label{sec:intro}

Image compression is an important step for enabling efficient transmission and storage of images in many applications. It is widely used in biomedical imaging due to the high-dimensional nature of data. While traditional image compression methods are based on fixed image transforms~\cite{Wallace1991, Rabbani.Joshi2002}, deep learning (DL) has recently emerged as a powerful data-driven alternative. The majority of DL-based compression methods are based on training autoencoders to be invertible mappings from image pixels to quantized latent representations~\cite{Cheng.etal2020,Minnen.etal2018,Lee.etal2018}. 

\medskip\noindent
In this work, we seek an alternative to the autoencoder-based compression methods by focusing on a recent paradigm using \emph{implicit neural representations (INRs)}. INR refers to a class of DL techniques that seek to learn a mapping from input coordinates (\emph{e.g.,} $(x,y)$) to the corresponding physical quantities (\emph{e.g.,} density at $(x,y)$) by using a \emph{multi-layer perceptron (MLP)}~\cite{Xie.etal2022, Mildenhall.etal2020, Shangguan.etal2022, Sun.etal2021, Reed.etal2021}. Recent studies have shown the potential of INR in image compression~\cite{Dupont.etal2021,Strumpler.etal2021,Dupont.etal2022,Ramirez.Gallego-Posada2022,Mancini.etal2022}.
The key idea behind INR based compression is to train a MLP to represent an image and consider the weights of the trained model as the compressed data. One can then reconstruct the image by evaluating the pre-trained MLP on the desired pixel locations. The traditional training strategy for image compression using INRs seeks to enforce image consistency between predicted and groundtruth image pixels. On the other hand, it is well-known that image quality can be improved by infusing prior knowledge on the desired images~\cite{Venkatakrishnan.etal2013,Romano.etal2017}. Based on this observation, we propose \emph{\textbf{S}tructural regular\textbf{I}zatio\textbf{N} for INR \textbf{CO}mpression (SINCO)} as new method for improving INR-based image compression using a new structural regularizer. Our structural regularizer seeks to improve the Dice score between the groundtruth segmentation maps and those obtained from the INR compressed image using a pre-trained segmentation network. We validate SINCO on brain MR images by showing that it can lead to significant improvements over the traditional INR-based image compression methods. We show that the combination of the traditional image-consistency loss and our structural regularizer enables SINCO to learn an INR that can better preserve desired image features. 

\begin{figure*}
    \centering
    \includegraphics[width=16.5cm]{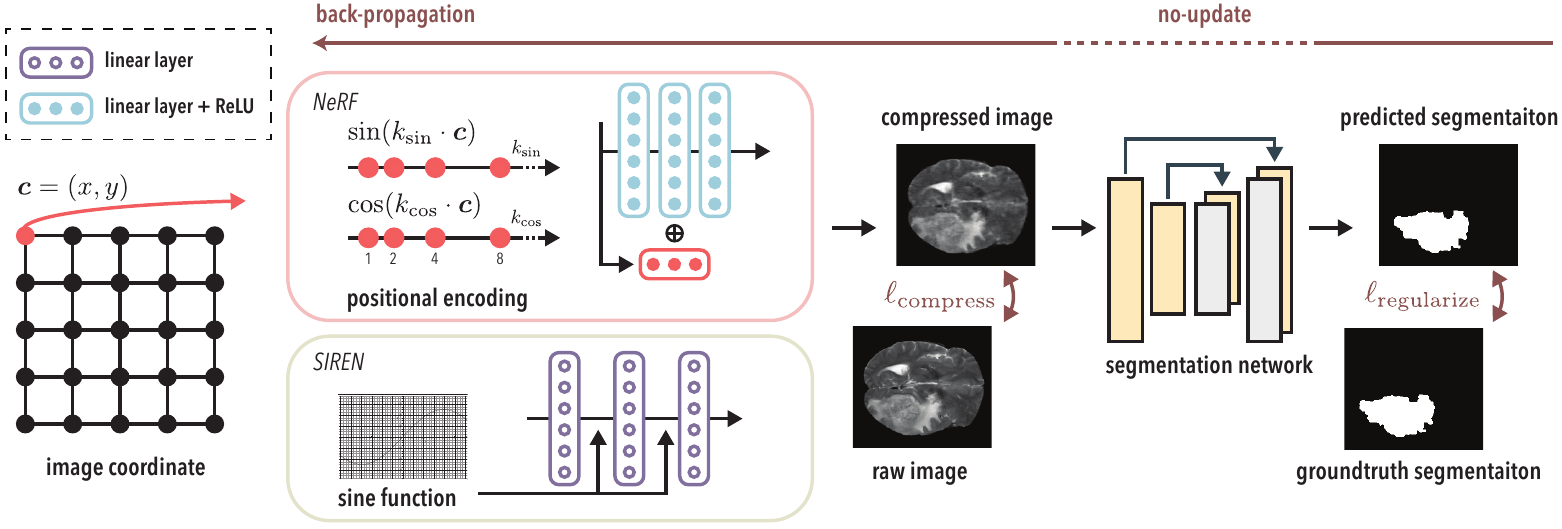} 
    \caption{SINCO consists of two components: (a) a multi-layer perceptron (MLP) for compressing an image by mapping its coordinates to the corresponding pixel values; (b) a segmentation network that predicts segmentation masks from the compressed image. Unlike traditional INR methods that only enforce consistency between compressed and groundtruth images, SINCO uses information from a regularizer that penalizes the discrepancy between predicted and groundtruth segmentation masks.
    }
    \label{fig:method}
\end{figure*}

\section{Background}
\label{sec:background}

INR (also referred to as \emph{neural fields}) denotes a class of algorithms for continuously representing physical quantities using coordinate-based MLPs (see a recent review~\cite{Xie.etal2022}). Recent work has shown the potential of INRs in many imaging and vision tasks, including novel view synthesis in 3D rendering~\cite{Mildenhall.etal2020}, video frame interpolation~\cite{Shangguan.etal2022}, computed tomography~\cite{Sun.etal2021} and dynamic imaging~\cite{Reed.etal2021}. The key idea behind INR is to train a MLP to map spatial coordinates to corresponding \emph{observed} physical quantities. After training, one can evaluate the pre-trained MLP on desired coordinates to predict the corresponding physical quantities, including on locations that were not part of training. Let $\cbm$ denotes a vector of input coordinates, $v$ the corresponding physical quantity, and $\mathcal{M}_\thetabm$ a MLP with trainable parameter $\thetabm\in\R^n$. The INR training can be formulated as
\begin{equation}
    \thetabmhat = \argmin_{\thetabm \in \R^n} \sum_{i=1}^N\ \ell_\mathsf{inr}(\mathcal{M}_\thetabm(\cbm_i), v_i)\ .
\end{equation}
where $N \geq 1$ denotes the number training pairs $(\cbm,v)$. The common choices for $\ell_\mathsf{inr}$ include $\ell_2$ and $\ell_1$ norms.

\medskip\noindent
INRs have been recently used for image compression~\cite{Dupont.etal2021,Strumpler.etal2021,Dupont.etal2022,Ramirez.Gallego-Posada2022} (see also a recent evaluation in medical imaging~\cite{Mancini.etal2022}). \emph{COmpressed Implicit Neural representations (COIN)}~\cite{Dupont.etal2021} is a pioneering work based on training a MLP by mapping the pixel locations (\emph{i.e., $\cbm=(x,y)$}) of an image to the pixel values. The pre-trained MLP in COIN is quantized and then used as the compressed data. In order to reconstruct the image, one can evaluate the model on the same pixel locations used for training. Several papers have investigated the meta-learning approach to accelerate COIN by first training a MLP over a large collection of datapoints and then fine-tuning it on an instance-dependent one~\cite{Dupont.etal2022,Strumpler.etal2021,Lee.etal2021}.
Two recent papers proposed to regularize INR-based image compression by using $\ell_0$- and $\ell_1$-norm penalties on the weights of the MLP to improve the compression rate~\cite{Strumpler.etal2021,Ramirez.Gallego-Posada2022}. 

\medskip\noindent
The structural regularization in SINCO is based on image segmentation using a pre-trained \emph{convolutional neural network (CNN)} (see a comprehensive review of the topic~\cite{Minaee.etal2021}). There exists a rich body of literature in the context of DL-based image segmentation that can be integrated into SINCO~\cite{Hesamian.etal2019,Isensee.etal2021,Ronneberger.etal2015}.
To the best of our knowledge, no prior work has considered higher-level structural regularization in the context of INR-based image compression.
It is worth mentioning that our structural regularizer is fully compatible to the existing INR compression methods; for example, one can easily combine our structural regularizer with an additional $\ell_0$-regularizer.

\section{Proposed Method}

SINCO consists of two components: \emph{(a)} a MLP $\mathcal{M}_\thetabm$ that represents an image by mapping its coordinates to corresponding pixel values and \emph{(b)} a CNN $\mathsf{g}_\varphi$ that predicts a segmentation mask given the compressed image produced by MLP. Specifically, let $\xbm\in\R^{H\times W}$ denote an image of height $H$ and width $W$ that we seek to compress. Let $\cbm\in\R^{HW\times 2}$ represents all the pixel locations within $\xbm$. Then, $\mathcal{M}_\thetabm$ is trained to take $\cbm$ as input and predict all the corresponding $HW$ pixels values. We format the output of $\mathcal{M}_\thetabm$ to be the compressed image $\xbmhat\in\R^{H\times W}$. The function $\mathsf{g}_\varphi$ denotes the segmentation CNN that takes the compressed image and predicts a segmentation mask $\hat{\sbm}=\mathsf{g}_\varphi(\xbmhat)$.
The SINCO pipeline is illustrated in Fig.~\ref{fig:method}.

\subsection{Network Architecture}
We implemented two different architectures for $\mathcal{M}_\thetabm$: 
\emph{(a)} SIREN\footnote{\url{http://github.com/lucidrains/siren-pytorch}.}~\cite{Sitzmann.etal2020} that
consists of linear layers followed by sine activation functions and \emph{(b)} NeRF that incorporates the positional encoding to expand $\cbm$ before passing it to $\mathcal{M}_\thetabm$\footnote{\url{http://github.com/wustl-cig/Cooridnate-based-Internal-Learning}}~\cite{Mildenhall.etal2020}
\begin{equation}
    \label{equ:pos_encoding}
    \gamma(\cbm) = \big(\sin(2^0\pi\cbm), \cos(2^0\pi\cbm) ... \sin(\underbrace{2^{L_f}\pi}_{k_{\sin}}\cbm), \cos(\underbrace{2^{L_f}\pi}_{k_{\cos}}\cbm)\big)
\end{equation}
where $L_f>0$ denotes the number of frequencies. $\mathcal{M}_\thetabm$ of NeRF consists of linear layers followed by ReLU activation functions. For the NeRF architecture, we also add residual connections from the input to intermediates layers.
We subsequently denote SINCO based on the two MLPs as \emph{SINCO (SIREN)} and \emph{SINCO (INR)}.
We adopt the widely-used U-Net~\cite{Ronneberger.etal2015} architecture as the CNN for the segmentation network $\mathsf{g}_\varphi$.


\subsection{Training Strategy}
SINCO is trained by minimizing the following loss
\begin{equation}
    \label{equ:loss}
    \ell_\mathsf{SINCO} = \ell_\mathsf{compress}(\xbmhat, \xbm) + \lambda\ell_\mathsf{regularize}(\hat{\sbm}, \sbm)\ ,
\end{equation}
where $\sbm$ denotes the reference segmentation mask of $\xbm$, and $\lambda \geq 0$ is a parameter that balances image consistency with structural regularization. We implement $\ell_\mathsf{compress}$ as a $\ell_2$-norm between the compressed image and the groundtruth image. We implement $\ell_\mathsf{regularize}$ as $1 - Dice(\hat{\sbm}, \sbm)$, where $Dice(\hat{\sbm}, \sbm)$ is the Dice score coefficent between the segmentation mask predicted from the compressed image and the groundtruth one. In eq.~\eqref{equ:loss}, the segmentation network $\mathsf{g}_\varphi$ is assumed to be pre-trained, which implies that we only optimize the parameters of $\mathcal{M}_\thetabm$ during training. Note that when $\lambda = 0$, eq.~\eqref{equ:loss} is equivalent to the traditional INR-based image compression. After the training, we follow the approach in~\cite{Dupont.etal2021} by quantizing the weights of $\mathcal{M}_\thetabm$ from 32-bits to 16-bits.

\begin{figure}
    \centering
    \includegraphics[width=8.5cm]{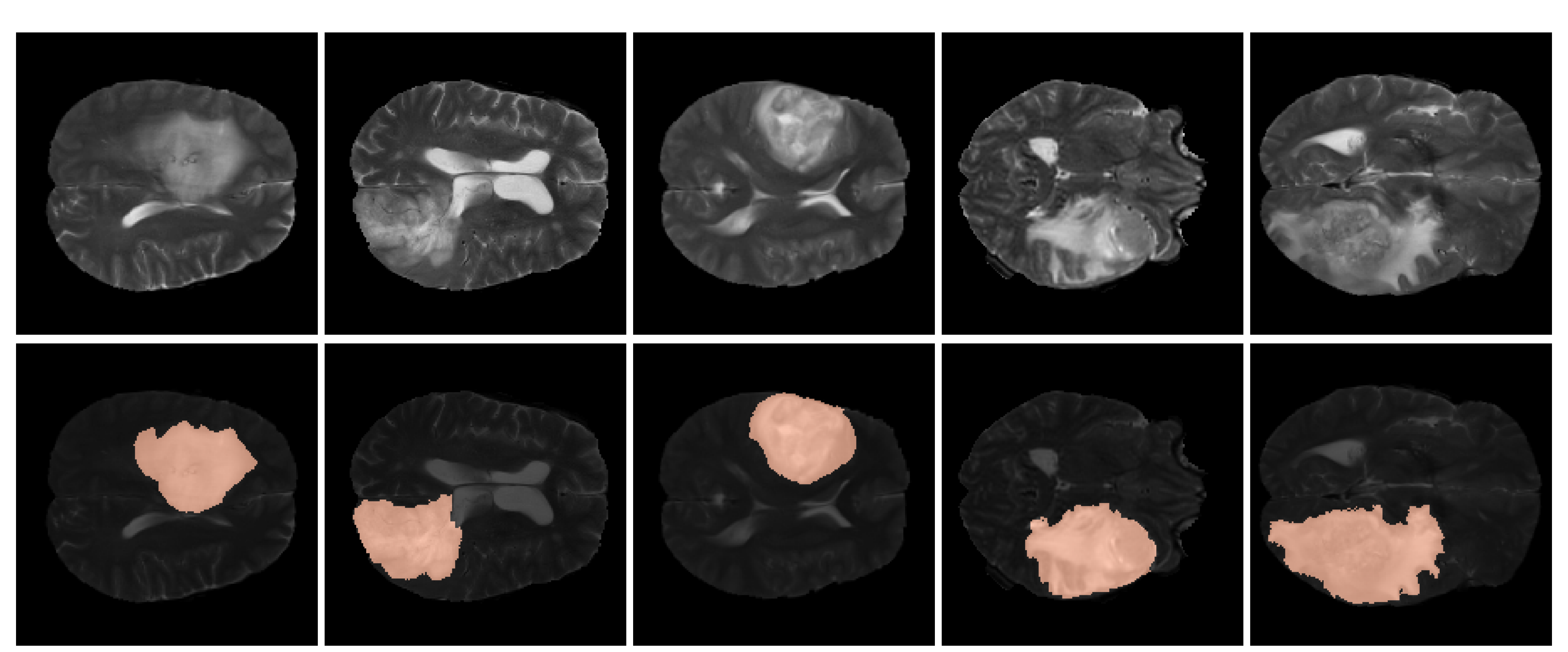}
    \caption{Test images (\emph{top}) and corresponding segmentation masks (\emph{bottom}) used for our numerical evaluations.
    }
    \label{fig:vis}
\end{figure}

\begin{figure*}
    \centering
    \includegraphics[width=16.5cm]{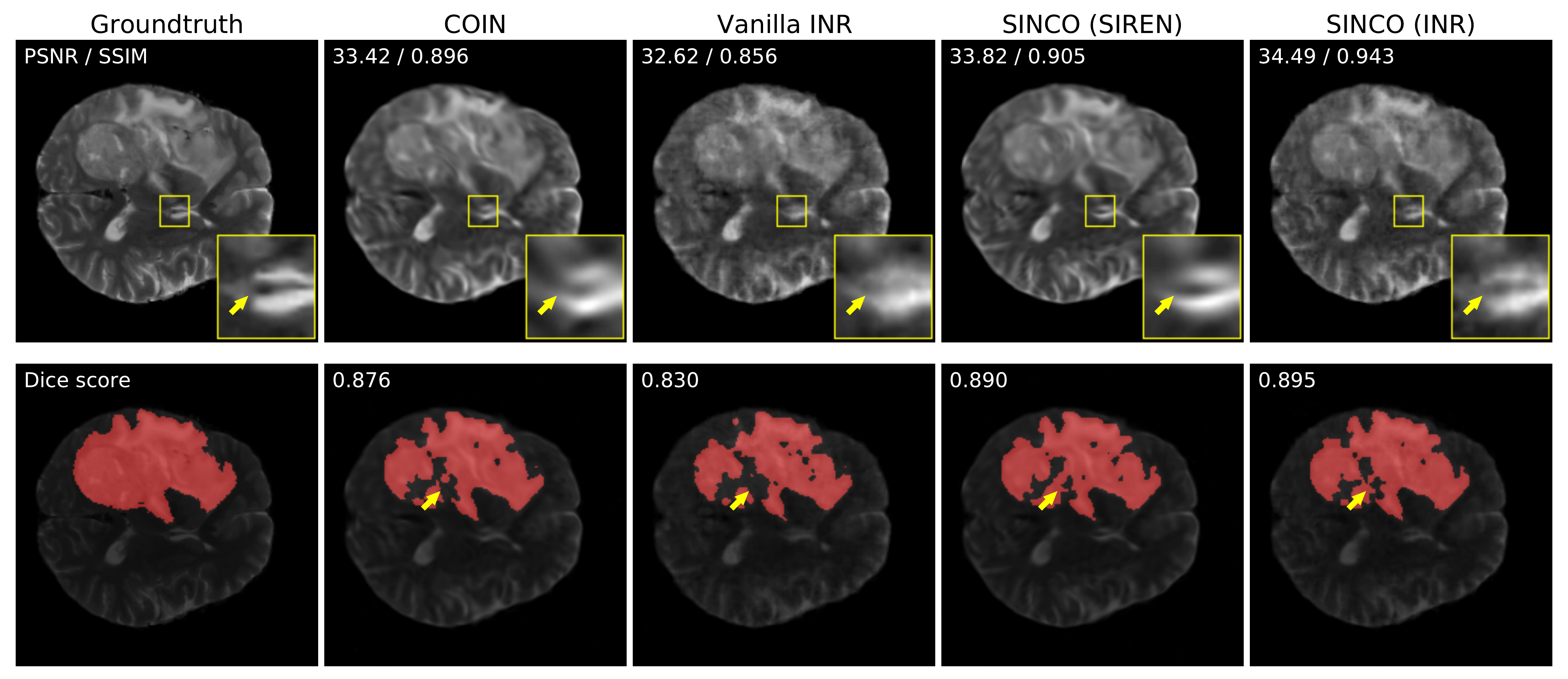}
    \caption{Visual illustration of compressed images (\emph{top}) with corresponding segmentation masks (\emph{bottom}) obtained by SINCO and two INR baselines. The PSNR and SSIM values for image compression and dice score values for image segmentation are provided in the top-left corners of the images. This figure shows that SINCO can lead to significant improvements in image quality both quantitatively and qualitatively.  Note the quality of the SINCO images in the areas highlighted by the yellow arrow.}
    \label{fig:result}
\end{figure*}

\begin{table}[t]
    \centering
    \small
    \renewcommand\arraystretch{1.1}
    \setlength{\tabcolsep}{3.7pt}

    \begin{tabular}{cccc}
    \toprule
    {Metrics}                   & PSNR (dB)       & SSIM & Dice score    \\
    \cmidrule{2-4}
    COIN               & 34.49 & 0.907 & 0.892  \\
    Vanilla INR  & 35.18 & 0.923 & 0.877  \\
    SINCO (SIREN) & 34.93 & 0.915 & 0.894  \\
    SINCO (INR)      & 35.69 & 0.952 & 0.895  \\
    \bottomrule 
    \end{tabular}
 
    \caption{Average quantitative evaluation metrics over the testing images. This table shows that SINCO can achieve better image compression and segmentation performance than INR baselines by imposing structural regularization.}

\label{tb:result}
\end{table}

\section{Numerical Experiments}

\subsection{Setup and comparison}

For our experiments, we used MR images of brain tumors with the corresponding segmentation masks obtained from the Decathlon dataset~\cite{Antonelli.etal2022} (Task01). We selected ten test images with resolution $240\times 240$. Fig.~\ref{fig:vis} illustrates several images used in our experiments with the corresponding segmentation masks for tumors.

\medskip\noindent
We compared SINCO against two reference INR methods: (a) \emph{COIN}\footnote{\url{http://github.com/EmilienDupont/coin}}, a recent INR method discussed in Sec.~\ref{sec:background}, 
and (b) \emph{Vanilla INR}, a variant of SINCO (INR) that sets $\lambda$ in \eqref{equ:loss} to $0$. Note that since COIN also uses SIREN as its MLP architecture, it can be viewed as SINCO (SIREN) using a different loss. For all methods we used the same compression rate quantified with \emph{bits per pixel (bpp)}
$$
    \mathsf{bpp} = \frac{\mathsf{bit\ per\ parameters}\times \mathsf{\#parameters}}{\mathsf{\#pixels}}\ .
$$
We set bbp to 1.2 in our experiments, corresponding to the compression rate of $2\%$ relative to the raw file size. We evaluated all the methods on both image compression and image segmentation using compressed images. For image compression, we used two widely-used quantitative metrics, \emph{peak signal-to-noise ratio (PSNR)}, measured in dB, and \emph{structural similarity index (SSIM)}. We used the \emph{Dice score} coefficient to evaluate image segmentation.

\subsection{Implementation Details} 

We followed~\cite{Ronneberger.etal2015} to train the segmentation network $\mathsf{g}_\varphibm$. We used about 500 images from the same dataset in training. The corresponding training loss can be written as
\begin{equation}
    \sum_{i=1}^M \ell_\mathsf{seg}(\mathsf{g}_\varphibm(\xbm_i), \sbm_i)
\end{equation}
where $M \geq 1$ denotes the number of training samples and $\ell_\text{seg}$ corresponds to binary cross entropy~\cite{Ronneberger.etal2015}. We used Adam~\cite{Kingma.Ba2017} as an optimizer with learning rate $0.0001$. We set training epochs to 75 and batch size to 8. In the training of SINCO, we set $\lambda$ in \eqref{equ:loss} to 1, and $L_f$ in \eqref{equ:pos_encoding} to 12. We used Adam~\cite{Kingma.Ba2017} as an optimizer with learning rate $0.001$. We set training epochs to 50,000. We performed our experiments on a machine equipped with an Intel Xeon Gold 6130 Processor and an NVIDIA GeForce RTX 1080Ti GPU.

\subsection{Results}
\begin{figure*}
    \centering
    \includegraphics[width=16.5cm]{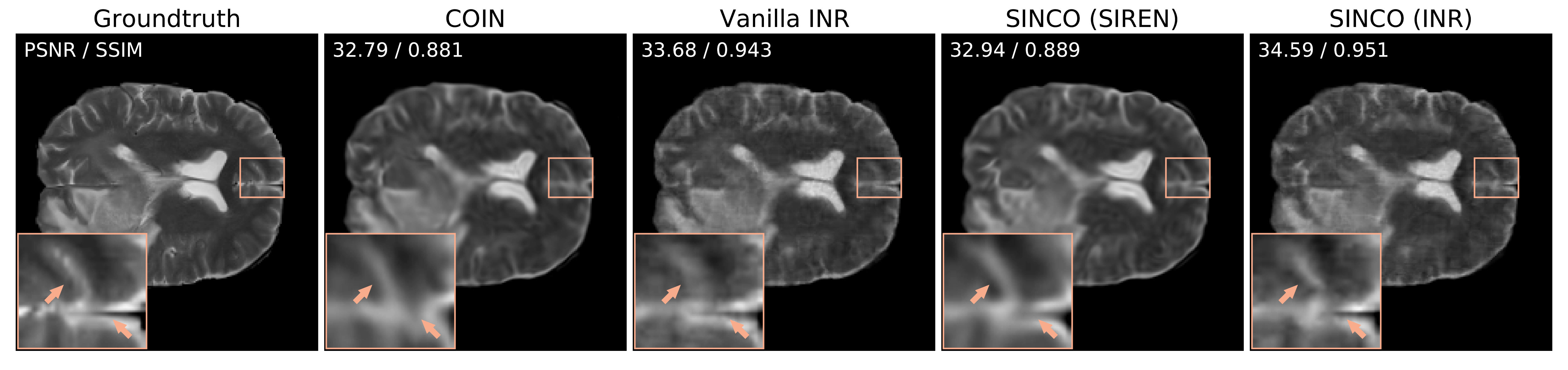}
    \caption{Visual illustration of compressed images obtained by SINCO and two other INR methods. The PSNR and SSIM values of compressed images relative to the groundtruth are provided in the top-left corner of the images. Note how SINCO provides images with better sharpness and overall quality by leveraging a structural regularizer (see features highlighted by arrows). }
    \label{fig:result_extra}
\end{figure*}

Figure~\ref{fig:result} presents visual results of SINCO and two other INR methods. For image compression, SINCO achieves higher PSNR and SSIM values than other INR methods, while also providing sharper images with features that are more consistent with the groundtruth.
For example, SINCO can reconstruct sharper brain tissue highlighted by the yellow arrow, while INR baselines blur out the same features.
For image segmentation using compressed images, SINCO can qualitatively and quantitatively outperforms INR baselines in general. Especially, SINCO (SIREN) provides more consistent results relative to the groundtruth segmentaion mask (see also segmentation mask highlighted by yellow arrows) compared with COIN. Note that COIN also uses SIREN as its MLP architecture. Another set of visual results shown in Figure~\ref{fig:result_extra} are also consistent with the observations that SINCO leads to significant improvements in the image quality. Table~\ref{tb:result} summarizes quantitative results over all the testing images. The table shows that, under the same MLP architecture (\emph{i.e.,} COIN versus SINCO (SIREN) and Vanilla INR versus SINCO (INR)), SINCO can achieve better performance by leveraging the structural regularization.

\section{Conclusion}

We present SINCO as a new structurally regularized image compression method using implicit neural representation. The key idea behind SINCO is to use a pre-trained segmentation network to ensure that the INR compressed images produce accurate segmentation masks. Our experiments on brain MR images show that SINCO can quantitatively and qualitatively outperform traditional INR approaches. In the future work, we will further investigate SINCO for higher dimensional data compression (\emph{e.g.,} 3D or 4D MRI~\cite{Eldeniz.etal2021}) and leverage recent development of meta-learning strategies to accelerate the INR training~\cite{Lee.etal2021}.

\section{Acknowledgements}
This material is based upon work supported by the NSF CAREER award under CCF-2043134. The authors are grateful to Yu Sun for helpful discussion.


\begin{thebibliography}{10}

\bibitem{Wallace1991}
G.~K Wallace,
\newblock ``The {{JPEG}} still picture compression standard,''
\newblock {\em Commun. ACM}, vol. 34, no. 4, pp. 30--44, 1991.

\bibitem{Rabbani.Joshi2002}
M.~Rabbani and R.~Joshi,
\newblock ``An overview of the {{JPEG}} 2000 still image compression
  standard,''
\newblock {\em Signal Process. Image Commun.}, vol. 17, no. 1, pp. 3--48, 2002.

\bibitem{Cheng.etal2020}
Z.~Cheng, H.~Sun, M.~Takeuchi, and J.~Katto,
\newblock ``Learned {{Image Compression With Discretized Gaussian Mixture
  Likelihoods}} and {{Attention Modules}},''
\newblock in {\em Proc. {{IEEE Conf}}. {{Comput}}. {{Vis}}. {{Pattern
  Recognit}}.}, 2020, pp. 7939--7948.

\bibitem{Minnen.etal2018}
D.~Minnen, J.~Ball{\'e}, and G.~D Toderici,
\newblock ``Joint {{Autoregressive}} and {{Hierarchical Priors}} for {{Learned
  Image Compression}},''
\newblock in {\em Proc. {{Adv}}. {{Neural Inf}}. {{Process}}. {{Syst}}.}, 2018,
  vol.~31.

\bibitem{Lee.etal2018}
J.~Lee, S.~Cho, and S.-K. Beack,
\newblock ``Context-adaptive entropy model for end-to-end optimized image
  compression,''
\newblock in {\em Proc. {{Int}}. {{Conf}}. {{Mach}}. {{Learn}}.}, 2018.

\bibitem{Xie.etal2022}
Y.~Xie \emph{et. al},
\newblock ``Neural fields in visual computing and beyond,''
\newblock {\em STAR}, vol. 41, no. 2, 2022.

\bibitem{Mildenhall.etal2020}
B.~Mildenhall, P.~P Srinivasan, M.~Tancik, J.~T Barron, R.~Ramamoorthi, and
  R.~Ng,
\newblock ``{NeRF}: {{Representing}} scenes as neural radiance fields for view
  synthesis,''
\newblock in {\em European {{Conf}}. on {{Comput}}. {{Vis}}.}, 2020, pp.
  405--421.

\bibitem{Shangguan.etal2022}
W.~Shangguan, Y.~Sun, W.~Gan, and U.~S. Kamilov,
\newblock ``Learning {{Cross-Video Neural Representations}} for {{High-Quality
  Frame Interpolation}},''
\newblock in {\em European {{Conf}}. on {{Comput}}. {{Vis}}.}, 2022.

\bibitem{Sun.etal2021}
Y.~Sun, J.~Liu, M.~Xie, B.~Wohlberg, and U.~S. Kamilov,
\newblock ``Coil: Coordinate-based internal learning for tomographic imaging,''
\newblock \emph{IEEE Trans. Comput. Imaging}, vol. 7, pp. 1400--1412, 2021.

\bibitem{Reed.etal2021}
A.~W.~Reed \emph{et. al},
\newblock ``Dynamic {{CT Reconstruction}} from {{Limited Views}} with
  {{Implicit Neural Representations}} and {{Parametric Motion Fields}},''
\newblock in {\em Proc. {{IEEE Int}}. {{Conf}}. {{Comput}}. {{Vis}}.}, 2021,
  pp. 2238--2248.

\bibitem{Dupont.etal2021}
E.~Dupont, A.~Goli{\'n}ski, M.~Alizadeh, Y.~W. Teh, and A.~Doucet,
\newblock ``{{COIN}}: {{COmpression}} with {{Implicit Neural}}
  representations,''
\newblock in {\em Proc. {{Int}}. {{Conf}}. {{Learn}}. {{Represent}}.
  {{Workshop}}}, 2021.

\bibitem{Strumpler.etal2021}
Y.~Str{\"u}mpler, J.~Postels, R.~Yang, L.~{van Gool}, and F.~Tombari,
\newblock ``Implicit {{Neural Representations}} for {{Image Compression}},''
\newblock {\em arXiv:2112.04267}, Dec. 2021.

\bibitem{Dupont.etal2022}
E.~Dupont, H.~Loya, M.~Alizadeh, A.~Goli{\'n}ski, Y.~W. Teh, and A.~Doucet,
\newblock ``{{COIN}}++: {{Neural Compression Across Modalities}},''
\newblock {\em arXiv:2201.12904}, June 2022.

\bibitem{Ramirez.Gallego-Posada2022}
J.~Ramirez and J.~{Gallego-Posada},
\newblock ``L0onie: {{Compressing COINs}} with {{L0-constraints}},''
\newblock {\em arXiv:2207.04144}, July 2022.

\bibitem{Mancini.etal2022}
M.~Mancini, D.~K. Jones, and M.~Palombo,
\newblock ``Lossy compression of multidimensional medical images using
  sinusoidal activation networks: An evaluation study,''
\newblock {\em arXiv:2208.01602}, Aug. 2022.

\bibitem{Venkatakrishnan.etal2013}
S.~V. Venkatakrishnan, C.~A. Bouman, and B.~Wohlberg,
\newblock ``Plug-and-{{Play}} priors for model based reconstruction,''
\newblock in {\em Proc. {{IEEE Glob}}. {{Conf}}. {{Signal Process}}. {{Inf}}.
  {{Process}}.}, 2013, pp. 945--948.

\bibitem{Romano.etal2017}
Y.~Romano, M.~Elad, and P.~Milanfar,
\newblock ``The {{Little Engine That Could}}: {{Regularization}} by
  {{Denoising}} ({{RED}}),''
\newblock {\em SIAM J. Imaging Sci.}, vol. 10, no. 4, pp. 1804--1844, 2017.

\bibitem{Lee.etal2021}
J.~Lee, J.~Tack, N.~Lee, and J.~Shin,
\newblock ``Meta-learning sparse implicit neural representations,''
\newblock in {\em Proc. {{Adv}}. {{Neural Inf}}. {{Process}}. {{Syst}}.}, 2021,
  vol.~34, pp. 11769--11780.

\bibitem{Minaee.etal2021}
S.~Minaee, Y.~Y Boykov, F.~Porikli, A.~J Plaza, N.~Kehtarnavaz, and
  D.~Terzopoulos,
\newblock ``Image segmentation using deep learning: {{A}} survey,''
\newblock {\em IEEE Trans. Pattern Anal. Mach. Intell.}, 2021.

\bibitem{Hesamian.etal2019}
M.~H. Hesamian, W.~Jia, X.~He, and P.~Kennedy,
\newblock ``Deep learning techniques for medical image segmentation:
  Achievements and challenges,''
\newblock {\em J. Digit. Imaging}, vol. 32, no. 4, pp. 582--596, 2019.

\bibitem{Isensee.etal2021}
F.~Isensee, P.~F Jaeger, S.~A. Kohl, J.~Petersen, and K.~H {Maier-Hein},
\newblock ``{{nnU-Net}}: A self-configuring method for deep learning-based
  biomedical image segmentation,''
\newblock {\em Nat. Methods}, vol. 18, no. 2, pp. 203--211, 2021.

\bibitem{Ronneberger.etal2015}
O.~Ronneberger, P.~Fischer, and T.~Brox,
\newblock ``U-{{Net}}: {{Convolutional}} networks for biomedical image
  segmentation,''
\newblock in {\em Proc. {{Medical Image Computing}} and {{Computer-Assisted
  Intervention}}}, 2015, pp. 234--241.

\bibitem{Sitzmann.etal2020}
V.~Sitzmann, J.~N.~P. Martel, A.~W. Bergman, D.~B. Lindell, and G.~Wetzstein,
\newblock ``Implicit {{Neural Representations}} with {{Periodic Activation
  Functions}},''
\newblock in {\em Proc. {{Adv}}. {{Neural Inf}}. {{Process}}. {{Syst}}.}, 2020,
  pp. 7462--7473.

\bibitem{Antonelli.etal2022}
M.~Antonelli \emph{et. al},
\newblock ``The {{Medical Segmentation Decathlon}},''
\newblock {\em Nat. Commun.}, vol. 13, no. 1, pp. 4128, July 2022.

\bibitem{Kingma.Ba2017}
D.~P. Kingma and J.~Ba,
\newblock ``Adam: {{A Method}} for {{Stochastic Optimization}},''
\newblock {\em arXiv:1412.6980}, Jan. 2017.

\bibitem{Eldeniz.etal2021}
C.~Eldeniz \emph{et. al},
\newblock ``{{Phase2Phase}}: {{Respiratory Motion-Resolved Reconstruction}} of
  {{Free-Breathing Magnetic Resonance Imaging Using Deep Learning Without}} a
  {{Ground Truth}} for {{Improved Liver Imaging}},''
\newblock {\em Invest. Radiol.}, vol. 56, no. 12, pp. 809--819, 2021.

\end{thebibliography}
\end{document}